\documentclass[aip,preprint,superscriptaddress,amsmath,amssymb,pof]{revtex4-1}
\usepackage{graphicx} 
\usepackage{dcolumn}
\usepackage{bm}
\usepackage{color}
\usepackage{float}
\usepackage{hyperref}
\hypersetup{
    colorlinks,%
    citecolor=blue,%
    filecolor=blue,%
    linkcolor=blue,%
    urlcolor=blue
} 

\begin{document}

\title{Modeling quasi-static magnetohydrodynamic turbulence with variable energy flux}

\author{Mahendra K. Verma}
\email[Email : ]{mkv@iitk.ac.in}
\affiliation{Department of Physics, Indian Institute of Technology -- Kanpur 208016, India}

\author{K. Sandeep Reddy}
\affiliation{Department of Mechanical Engineering, Indian Institute of Technology -- Kanpur 208016, India}

\date{\today}

\begin{abstract}
In quasi-static MHD, experiments and numerical simulations reveal that the energy spectrum is steeper than  Kolmogorov's  $k^{-5/3}$ spectrum.  To explain this observation, we  construct turbulence models based on variable energy flux, which is caused by the Joule dissipation.  In the first model, which is applicable to small interaction parameters, the energy spectrum is a power law, but with a spectral exponent steeper than -5/3. In the other limit of large interaction parameters, the second model predicts an exponential energy spectrum and flux. The model predictions are in good agreement with the numerical results.
\end{abstract}


\maketitle

\section{Introduction}\label{sec:int}
The liquid metal flows  in fission and fusion reactors, and metal plate rolling and crystallization have very small magnetic  Reynolds number $\mathrm{Rm} = U L/\eta$, where  $U,L$ are the large scale velocity and length scales respectively, and $\eta$ is the magnetic diffusivity.  In this paper, we will construct several models to derive energy spectrum and flux for an idealized limit, called the ``quasi-static limit", for which $\mathrm{Rm}\rightarrow 0$.

In the quasi-static limit, the induced magnetic field tends to be very small  because of very  large magnetic diffusivity, and it gets slaved to the velocity field that yields the Lorentz force as
\begin{equation}
{\bf F} = -\frac{\sigma B_0^2}{\rho} \frac{1}{\nabla^2} \frac{\partial^2 {\bf u}}{\partial z^2},
\end{equation}
where $\rho$ is the density of the fluid, ${\bf u}$ is the velocity field, and ${\bf B} = B_0 \hat{z}$ is the external uniform magnetic field.\cite{Davidson:book,Moreau:book}   The quasi-static approximation provides a major simplification since  we do not need to solve the induction equation.  The strengths of the Lorentz force and the external magnetic field are quantified using a nondimensionalized parameter called the  ``interaction parameter", which is a ratio of the Lorentz force and the nonlinear term.   

Several experimental and numerical simulations have been performed to study energy spectrum of quasi-static MHD turbulence (see Knaepen and Moreau,\cite{Knaepen:ARFM2008} and references therein).  Kolesnikov and Tsinober,\cite{Kolesnikov:FD1974} and Alemany {\it et al.}\cite{Alemany:JMec1979} performed experiments on mercury for low $\mathrm{Rm}$, and observed that the energy spectrum for the velocity field follows $k^{-3}$ scaling for significantly strong interaction parameters.   A similar experiment by Branover {\it et al.}\cite{Branover:PTR1994}  on mercury showed energy spectrum -- $k^{-5/3},k^{-7/3},k^{-3},k^{-11/3}$ --  for different interaction parameters; the exponents below $-3$ were attributed to  the generation of helicity in the flows.   In an experiment on liquid sodium, Eckert {\it et al.}\cite{Eckert:HFF2001} observed the energy spectrum  to follow $k^{-\alpha}$, where $\alpha \in [5/3,5]$ for interaction parameter $N \in [0.3, 1000]$. 


Many numerical simulations of the quasi-static MHD~\cite{Reddy:POF2014,Zikanov:JFM1998,Vorobev:POF2005,Ishida:POF2007,
Knaepen:ARFM2008,Favier:POF2010b,Favier:JFM2011,Burattini:PD2008,Burattini:PF2008} show steepening of the energy spectrum with the increase of interaction parameter, similar to those seen in the experiments. It has been observed that for large interaction parameters, the flow becomes  anisotropic with the energy concentrated near the plane perpendicular to the external magnetic field.\cite{Caperan:JDM1985,Potherat:JFM2010,Reddy:POF2014,Zikanov:JFM1998,Burattini:PF2008,Favier:POF2010b} 
 Recently, Reddy and Verma\cite{Reddy:POF2014} performed simulations for interaction parameters ranging from 0 to 220, and showed that the energy spectrum is power law for $0 < N < 27$, and exponential ($\exp(-b k)$) for $N \ge 130$.  Ishida and Kaneda\cite{Ishida:POF2007} studied the modification of inertial range energy spectrum for low interaction parameters, and proposed a $k^{-7/3}$ scaling law. Burattini {\it et al.}\cite{Burattini:PD2008} studied anisotropy in quasi-static MHD turbulence and also observed a scaling law different from  $k^{-3}$ for the energy spectrum. 
 
To understand the numerical and experimental findings, in this paper we construct turbulence models for quasi-static MHD turbulence.  Our model is based on the fact that the energy flux decreases with wavenumber due to the Joule dissipation.\cite{Reddy:ARXIV2014}
For small interaction parameters, the turbulence is still isotropic to a large extent;  our model provides the energy spectrum and energy flux for a given interaction parameter.  For  large interaction parameters, however, the spectrum is highly anisotropic and has an exponential dependence on $k$.  We  derive the energy flux and spectrum for this regime using variable energy formalism.   We show that our model results are  consistent with the earlier numerical \cite{Reddy:POF2014,Burattini:PD2008} and experimental results.\cite{Branover:PTR1994,Eckert:HFF2001}  We also perform numerical simulations to validate our models.   We remark that similar steepening of energy spectrum was observed by Verma\cite{Verma:EPL2012} in two-dimensional turbulence with Ekman friction.

The organization of the paper is as follows.  In Sec.~\ref{sec:theory} we describe the variable energy flux models for quasi-static MHD.    Validation of the models using numerical simulations are discussed in Sec.~\ref{sec:results}.  Section~\ref{sec:conclusions} contains conclusions.

\section{Theoretical Framework}\label{sec:theory}
The governing equations for the low-Rm liquid metal flows under the quasi-static approximation are\cite{Davidson:book,Moreau:book}
\begin{eqnarray}
\frac{\partial {\bf u}}{\partial t} + {\bf u \cdot \nabla u}  &=& -\nabla (p/\rho) 
	 -\frac{\sigma B_0^2}{\rho} \frac{1}{\nabla^2} \frac{\partial^2 {\bf u}}{\partial z^2} + \nu \nabla^2 {\bf u}, \label{eq:NS1} \\
\nabla \cdot {\bf u} &=& 0, \label{eq:incomp}
\end{eqnarray}
where ${\bf u}$ is the velocity field, $p$ is the pressure field, ${\bf B}_0$ is the uniform external magnetic field along the $z$ direction, $\sigma$ is the electrical conductivity, $\nu$ is  the kinematic viscosity, and $\rho$ is the density of the fluid.  The corresponding equation in the Fourier space,
\begin{equation}
\frac{\partial \hat u_i ({\bf k)}} {\partial t} = -i  k_i \frac{\hat p({\bf k})}{\rho} - i k_j \sum \hat u_j ({\bf q}) \hat u_i( {\bf k-q})
	-\frac{\sigma B_0^2}{\rho} (\cos^2 \theta) \hat u_i({\bf k}) - \nu k^2  \hat u_i ({\bf k}),
\end{equation}
is very useful in analyzing energy transfers among modes.  Here $\theta$ is the angle between the mean magnetic field and the wavenumber ${\bf k}$ (see Fig.~\ref{fig:ring_decomposition}).   We define interaction parameter $N$ as the ratio of the Lorentz force and the nonlinear term:
\begin{equation}
N = \frac{\sigma B_0^2 L}{\rho U}.
\end{equation}    
For large external magnetic field, $N$ is large, and flow is strongly anisotropic.
  
    The  energy equation in the Fourier space is\cite{Davidson:book,Moreau:book}
\begin{equation}
\frac{\partial E ({\bf k})} {\partial t}   = T ({\bf k}) - 2 \frac{\sigma B_0^2}{\rho} \cos^2 (\theta) E ({\bf k}) - 2 \nu k^2  E ({\bf k}),
\label{eq:energy}
\end{equation}
where $E ({\bf k}) = |{\bf \hat u(k)}|^2/2$ is the energy spectrum, and $T({\bf k})$ is the kinetic energy transfer rate.   The second and   third terms in the RHS are the dissipation rates due to the Lorentz force and the viscous force respectively.
 
 \subsection{Variable energy flux}

For zero interaction parameter, which is the fluid (hydrodynamic) limit,  the flow becomes turbulent when  Reynolds number $\mathrm{Re}=UL/\nu \gg 1$.  In this regime, the energy spectrum exhibits the famous Kolmogorov's $k^{-5/3}$ power law in the inertial range.   For finite $N$, however, the Lorentz force induces an additional dissipation that leads to a modification of the energy flux.  The variation of the energy flux due to this dissipation can be derived using the following arguments.  
 
\begin{figure}[htbp]
 \begin{center}
 \includegraphics[scale=0.40]{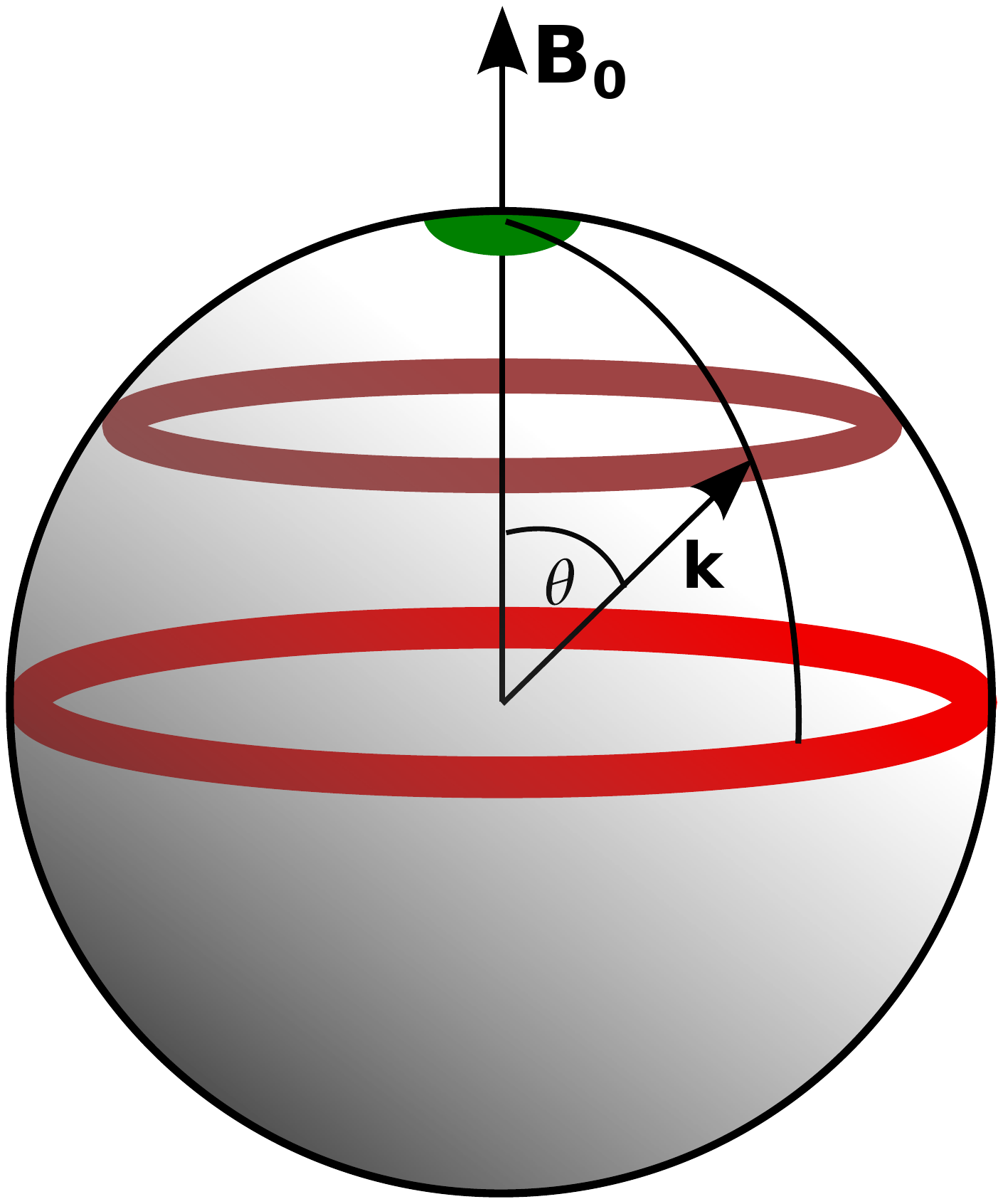}
 \end{center}
\caption{Figure illustrating ring decomposition in spectral space.}  
 \label{fig:ring_decomposition}
 \end{figure}

We assume that the energy spectrum is anisotropic due to the mean magnetic field,\cite{Burattini:PD2008,Potherat:JFM2010,Reddy:POF2014} and it is described using the  ring spectrum $E(k,\theta)$,\cite{Teaca:PRE2009,Burattini:PD2008}  where $k$ is the wavenumber of the ring, and $\theta$ is the angle between the mean magnetic field and the ``average'' wavenumber ${\bf k}$ of the ring, as shown in  
Fig.~\ref{fig:ring_decomposition}. 
We model $E(k,\theta)$ as
 \begin{equation}
E(k,\theta) = E(k) \frac{g(\theta)}{\pi},
\label{eq:Ek}
\end{equation}
where $g(\theta)$ describes the angular dependence of the energy spectrum.  An integration of Eq.~(\ref{eq:Ek}) over $\theta$ yields
\begin{equation}
\int_0^{\pi}  d\theta E(k,\theta) = E(k) \int_0^{\pi}  \frac{g(\theta)}{\pi} = E(k).
\end{equation}
Therefore, 
\begin{equation}
  \int_0^{\pi}  \frac{g(\theta)}{\pi} = 1.
\end{equation}
 For the isotropic case, $g(\theta) = \mathrm{const}=1$.

Due to  Joule dissipation, the inertial-range energy flux $\Pi(k)$ decreases  with the increase of $k$.  Quantitively, the difference between energy fluxes $\Pi(k+dk)$ and $\Pi(k)$ is due to the energy dissipation in the shell $(k,k+dk)$, i.e., 
 \begin{equation}
 \Pi(k+dk) - \Pi(k) = - \epsilon(k) dk = -\left\{ \int_0^\pi d\theta \left[ 2 \nu  k^2 + 2 \frac{\sigma B_0^2}{\rho} \cos^2 \theta \right] E(k,\theta) \right\} dk,
\end{equation}
or
 \begin{equation}
 \frac{d\Pi(k)}{dk} = -\left[ 2c_1  \nu k^2 +2c_2 \frac{\sigma B_0^2}{\rho}  \right] E(k),
 \label{eq:dPidk}
\end{equation}
with 
 \begin{eqnarray}
c_1 & = &\frac{1}{\pi} \int_0^\pi g(\theta) d \theta = 1 \label{eq:c1} \\
c_2 & = &  \frac{1}{\pi} \int_0^\pi g(\theta) \cos^2 \theta d \theta. \label{eq:c2}
\end{eqnarray}

In the following discussion, we will construct two models: model $A$ for small $N$'s for which the energy spectrum is still a power law but steeper than Kolomogorov's $k^{-5/3}$ spectrum; and model $B$ for large $N$ for which the energy spectrum is exponential.   The energy spectra and fluxes for the two cases are derived self-consistently using Eq.~(\ref{eq:dPidk}).

\subsection{Model $A$ for small interaction parameters}
\label{sec:modelA}
In the present subsection, we describe a formalism of variable energy flux for small and moderate interaction parameters.  Motivated by the experimental and simulation results, for this range of $N$, we postulate a power law for the energy spectrum.   Specifically, we extrapolate Pope's shell spectrum \cite{Pope:book} for the isotropic turbulence to the ring spectrum as
\begin{equation}
E(k,\theta) = E(k)  \frac{g(\theta)}{\pi} = C [\Pi(k)]^{2/3} k^{-5/3}  f_L(k L) f_\eta(k \eta) \frac{g(\theta)}{\pi},
\label{eq:Ek_smallN}
\end{equation}
where $C$ is the Kolmogorov's constant with an approximate value of 1.5, $\Pi(k)$ is the energy flux emanating  from the wavenumber sphere of radius $k$, and $g(\theta)$ is the anisotropic component of the energy spectrum. The functions $f_L(kL)$ and $f_\eta(k \eta)$ specify the  large-scale and dissipative-scale components, respectively, of the energy spectrum:
\begin{eqnarray}
f_L(kL) & = & \left( \frac{kL}{[(kL)^2 + c_L]^{1/2}} \right)^{5/3+p_0}, 
\label{eq:fL} \\
f_\eta(k\eta) & = & \exp \left[ -\beta \left\{ [ (k\eta)^4 + c_\eta^4 ]^{1/4}   - c_\eta \right\} \right],
\label{eq:feta}
\end{eqnarray}
where the $c_L, c_\eta, p_0, \beta$ are constants.  We take  $C_L \approx 6.78$, $c_\eta \approx 0.40$, $\beta \approx 5.2$, and $p_0 =2$, as suggested by Pope.\cite{Pope:book}  
In the present paper we focus on the inertial and dissipative range, hence, $f_L(kL) = 1$.

We substitute the energy spectrum of the form Eq.~(\ref{eq:Ek_smallN}) in Eq.~(\ref{eq:dPidk}), which yields
 \begin{equation}
 \frac{d\Pi(k)}{dk} = -\left[ 2c_1  \nu k^2 +2c_2 \frac{\sigma B_0^2}{\rho}  \right] C (\Pi(k))^{2/3} k^{-5/3} f_\eta(k \eta).
 \label{eq:dPidk_smallN}
\end{equation}
We integrate Eq.~(\ref{eq:dPidk_smallN}) from $k=k_1$, which is the starting wavenumber of the inertial range.  Assuming that the energy flux at this wavenumber is $\Pi_0$, we obtain
\begin{eqnarray}
\left[ \frac{\Pi(k)}{\Pi_0} \right]^{1/3} & = & 1- \frac{2C c_1}{3} 
			\left( \frac{\nu^3}{\Pi_0 \eta^4} \right)^{1/3} I_1(k \eta)
			-\frac{2 c_2 C \sigma B_0^2}{3\rho} \frac{\eta^{2/3}}{\Pi_0^{1/3}} I_2(k\eta) 
			\nonumber \\
		& = & 1-\frac{2c_1 c_3 C  }{3}   I_1(k \eta) - \frac{2}{3}\frac{c_2 C N}{\sqrt{c_3 Re}}   I_2(k\eta),
		\label{eq:flux_exact}
\end{eqnarray}
 where $\eta$ is the Kolmogorov length,  the dimensionless constant $c_3 = (\nu^3/\Pi_0 \eta^4)^{1/3}$, and the integrals $I_1$ and $I_2$ are
\begin{eqnarray}
I_1(k \eta) & = & \int_{k_1 \eta}^{k \eta} dk' k'^{1/3} f_\eta(k' ), \\
I_2(k \eta) & = & \int_{k_1 \eta}^{k \eta}  dk' k'^{-5/3} f_\eta(k').
\end{eqnarray}
We choose $c_3 = 3.1$ in order to achieve $\Pi(k) \rightarrow 0$ for $k \eta \gg 1$ when $N=0$ (isotropic case), and Kolmogorov's constant $C= 1.5$.  We also take 
 \begin{eqnarray}
c_1 & = &1  \label{eq:c1_smallN}  \\
c_2 & = & 1/2, \label{eq:c2_smallN}
\end{eqnarray}
which are the values  when $g(\theta)= \mathrm{const} = 1$, the isotropic case.  

To compare the aforementioned model with simulations, in which we force the wavenumbers $1\le |{\bf k}| \le 3$, we assume that the  inertial range wavenumber starts at around $k_1 = 4 \times 2 \pi/L$.   Therefore, the lower limit of the integral is $k_1 \eta = 4 (2 \pi) (\eta/L) = 8 \pi \times  (c_3 Re)^{-3/4} $.   Note that the energy flux $\Pi(k)$ peaks at  $k=k_1$ with value $\Pi_0$.

Equation (\ref{eq:flux_exact}) indicates that the second term, which arises due to the Lorentz force, is proportional to $N$.  Hence, the flux decreases significantly as $N$ is increased.  The form of $\Pi(k)$ can be derived in the  limiting case $\nu \rightarrow 0$, for which,  in the inertial range 
\begin{equation}
\left[ \frac{\Pi(k)}{\Pi_0} \right]^{1/3} \approx 1- \frac{c_2 C N}{\sqrt{c_3 Re}}  \left[ (k_1 \eta)^{-2/3} - (k \eta)^{-2/3} \right].
\end{equation}
Thus, $\Pi(k)$ decreases with an increase of $N$.


We compute the energy spectrum using the aforementioned $\Pi(k)$:
\begin{equation}
E(k) =
\begin{cases} 
C \Pi_0^{2/3} k^{-5/3}   f_\eta(k \eta) \left[ \frac{\Pi(k)}{\Pi_0} \right]^{2/3}, & \text{if  $k>k_1,$}  \\
C \Pi_0^{2/3} k^{-5/3}  f_L(k L), & \text{otherwise.}
\label{eq:spectrum_exact}
\end{cases}
\end{equation}
Thus, our model predicts a variable energy flux and a steeper energy spectrum, yet a power law spectrum.   In Sec.~\ref{subsec:small}, we will compare these predictions with numerical results.  

For  interaction parameter far above unity, the turbulence tends be strongly anisotropic, and the energy spectrum tends to deviate strongly from  Eqs.~(\ref{eq:Ek_smallN}).  These features make the above formalism inapplicable to $N > 1$.  Note that for $N \gg 1$, the energy spectrum is  exponential,  rather than a power law.\cite{Reddy:POF2014} It is very difficult and cumbersome to derive a general formalism for an arbitrary $N$, however, it is quite easy to derive a model for a very large interaction parameter, that will be described in the following subsection.

\subsection{Model $B$ for a very large interaction parameter}\label{subsec:modelB}
\label{sec:modelB}
In model $A$ described in the earlier subsection, we assume the energy spectrum to be a power law in $k$ (see Eq.~(\ref{eq:Ek_smallN})).  Numerical simulations and experiments show that this approximation is valid only for small and moderate $N$.   For very large $N$, the increase of the Joule dissipation on all scales causes a rapid decrease of  energy flux in the inertial range, resulting in an exponential behavior of energy spectrum.\cite{Reddy:POF2014} Therefore, for very large $N$, it is best to take an exponential form for  the energy flux, energy spectrum, and dissipation spectrum $\epsilon(k)$ since they satisfy Eq.~(\ref{eq:dPidk}).

For $N \gg 1$, we postulate that the energy spectrum and dissipation spectrum $\epsilon(k)$ follow
\begin{eqnarray}
E(k) & = & A \exp(-b k), \label{eq:Ek_largeN} \\
\epsilon(k) & = & \frac{d\Pi(k)}{dk} =  (P k^2 + Q) \exp(-b k), \label{eq:epsilonk_largeN}
\end{eqnarray}
where $A, P, Q$, and  $b$ are parameters, and $E(k,\theta)= E(k) g(\theta)/\pi$.   An integration of Eq.~(\ref{eq:epsilonk_largeN}) yields
\begin{equation}
\Pi(k)  =  \left\{ P \left(\frac{k^2}{b}+\frac{2k}{b^2}+\frac{2}{b^3}\right) + \frac{Q}{b} \right\} \exp(-b k).
\label{eq:Pik_largeN}
\end{equation}
A comparison of Eq.~(\ref{eq:epsilonk_largeN}) with Eq.~(\ref{eq:dPidk}) yields
\begin{eqnarray}
P &=& 2Ac_1 \nu, \label{eq:largeN_P}\\
Q &=& 2Ac_2 \frac{\sigma B_0^2}{\rho}. 
\label{eq:largeN_Q}
\end{eqnarray}
Thus, we show that the exponential energy spectrum and flux are consistent solutions of the variable flux equation (Eq.~(\ref{eq:dPidk})).  In Sec.~\ref{subsec:large}, we verify the above predictions with numerical simulations.

We performed numerical simulations to verify the model predictions described in this
section.  The simulation details and results will be described in the next  section.
 
\section{Validation of the Models Using Numerical Simulations }\label{sec:results} 
We simulate the quasi-static MHD using pseudo-spectral method.  We nondimensionalize  Eqs.~(\ref{eq:NS1},\ref{eq:incomp}) using the characteristic velocity $U_0$ as the velocity scale, the box dimension $L_0$ as the length scale, and $L_0/U_0$ as the time scale, and obtain
\begin{eqnarray}
\dfrac{\partial{\bf U}}{\partial T} + ({\bf U}\cdot\nabla'){\bf U} &=& -\nabla'{P} - B^{\prime 2}_0 \dfrac{1}{\nabla^{\prime2}} \dfrac{\partial^2{\bf U}}{\partial Z^2} + \nu^\prime \nabla^{\prime 2} {\bf U} + {\bf f^\prime}, \label{eq:NS2}  \\
\nabla^\prime \cdot {\bf U} & = & 0, \label{eq:continuity2} 
\end{eqnarray}
where non-dimensional variables are: $\mathbf U = \mathbf u/U_0$, $ \nabla' = L_0 \nabla $, $T = t(U_0/L_0)$, 
$B_0^{\prime 2}  = \sigma B_0^2 L_0 /(\rho U_0)$, and $\nu^\prime=\nu/(U_0 L_0)$. 

We use pseudo-spectral code {\it{Tarang}}\cite{Verma:Pramana2013} to solve the non-dimensional  Eqs.~(\ref{eq:NS2},\ref{eq:continuity2}) in a cube  with $256^3$ and $512^3$ grids, and with periodic boundary conditions applied in all the three directions. We use fourth-order Runge-Kutta method for time-stepping,  Courant-Friedrichs-Lewy (CFL) condition for calculating time-step ($\Delta t$), and the $3/2$ rule for dealiasing. In order to achieve a steady-state, the velocity field is randomly forced in the wavenumber band   $1 \leq {\bf |k|} \leq 3 $.


\begin{table}[htbp]
\caption{\label{tab:table}Table depicting various parameters used: the grid size, non-dimensional magnetic field $B'_0$, the Reynolds number $\mathrm{Re}$, the interaction parameter $N$ calculated at the steady state, the  energy spectrum, and non-dimensional viscosity $\nu'$.  }
\begin{ruledtabular}
\begin{tabular}{cccccc}
Grid 	& $B'_0$	& $\mathrm{Re}$		& $N$		& scaling law	& $\nu'$\\ 
\hline
$512^3$	& 0		& 480		& 0			& $k^{-5/3}$		& 0.00016 \\
$512^3$	& 0.739	& 460		& 0.10		& $k^{-1.8}$		& 0.00016 	\\
$512^3$	& 1.65	& 440   		& 0.64		& $k^{-2.0}$		& 0.00016 \\
$512^3$	& 2.34	& 370    	& 1.6		& $k^{-2.8}$		& 0.00016 \\ 
$256^3$	& 25.1	& 430		& 130		& $\mathrm{exp}(-0.18k)$		& 0.00036 \\
$256^3$	& 32.6	& 440		& 220		& $\mathrm{exp}(-0.18k)$		& 0.00036 \\
\end{tabular}
\end{ruledtabular}
\end{table}

We  simulated quasi-static MHD for interaction parameters $N= 0.1, 0.64$, and 1.6, belonging to small $N$ regime, and for $N=130$ and 220, belonging to the very large $N$ limit. The final state of a fluid run was used as the initial condition for the above $N$ runs.  All the simulations were carried out till a statistical steady state is reached.  The interaction parameter $N$ for each run was computed using
\begin{equation}
N = \frac{B_0'^2 L}{U'},
\end{equation} 
were $U'$ is the root mean square (rms) of the steady-state velocity, and $L$ is the non-dimensional integral length scale at the steady state. The Reynolds number is defined as 
\begin{equation}
\mathrm{Re} = \frac{U'L}{\nu^\prime}.
\end{equation}
 For further details on numerical simulations, refer to Reddy and Verma.\cite{Reddy:POF2014}


We compute the energy spectra and fluxes for $N = 0.10, 0.64, 1.6$ (small), as well as for $N=130,220$ (large).  We compare these numerical results with  model predictions.

\subsection{Small interaction parameters}\label{subsec:small}
\begin{figure}[h]
\begin{center}
\includegraphics[scale=0.5]{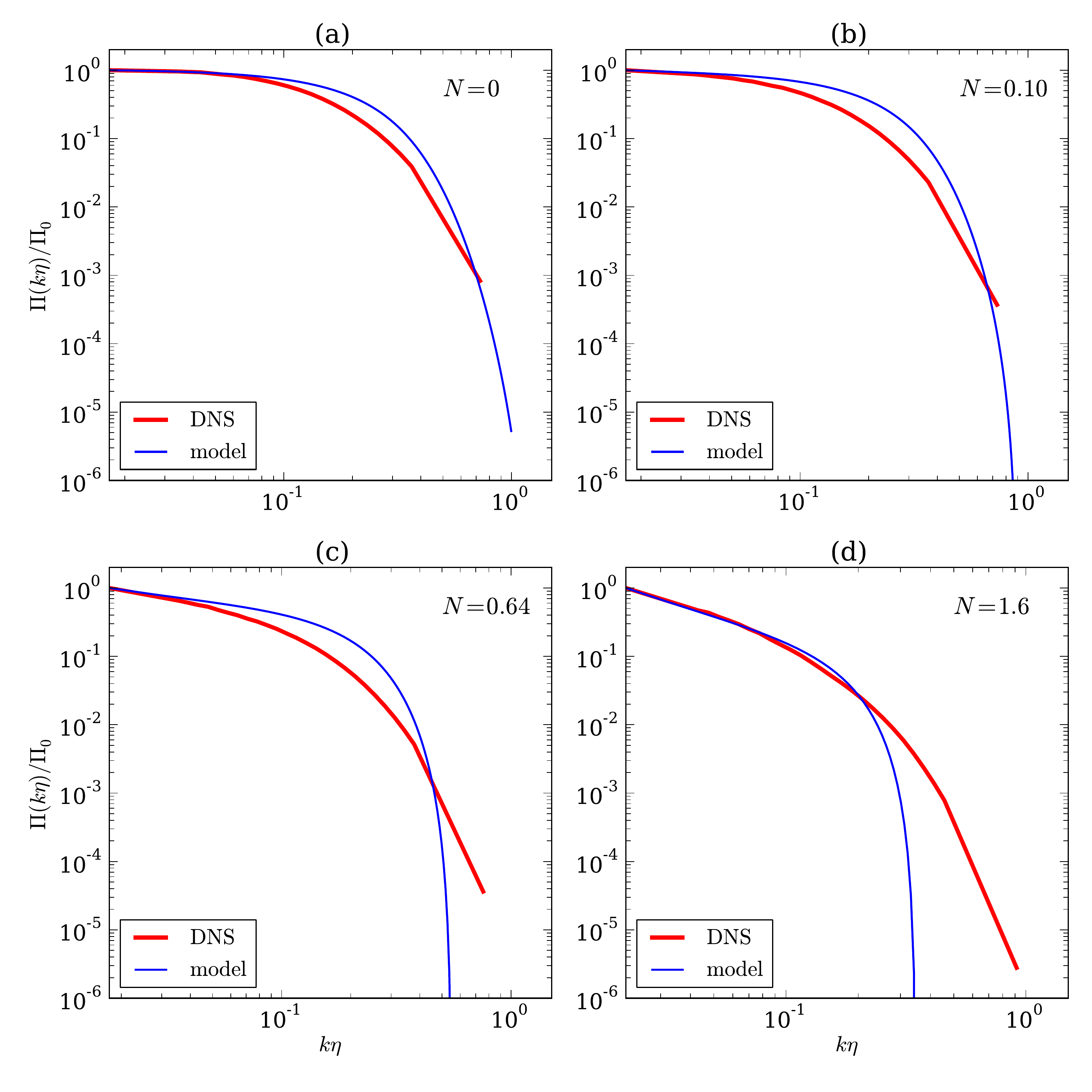}
\caption{ Plots of normalized energy flux $\Pi(k\eta)/\Pi_0$ for: (a) $N=0$, (b) $N=0.10$, (c) $N=0.64$, and (d) $N=1.6$. In all the cases, the energy flux decrease with $k$ due to Joule dissipation.}
\label{fig:flux}
\end{center}
\end{figure}

\begin{figure}[h]
\begin{center}
\includegraphics[scale=0.5]{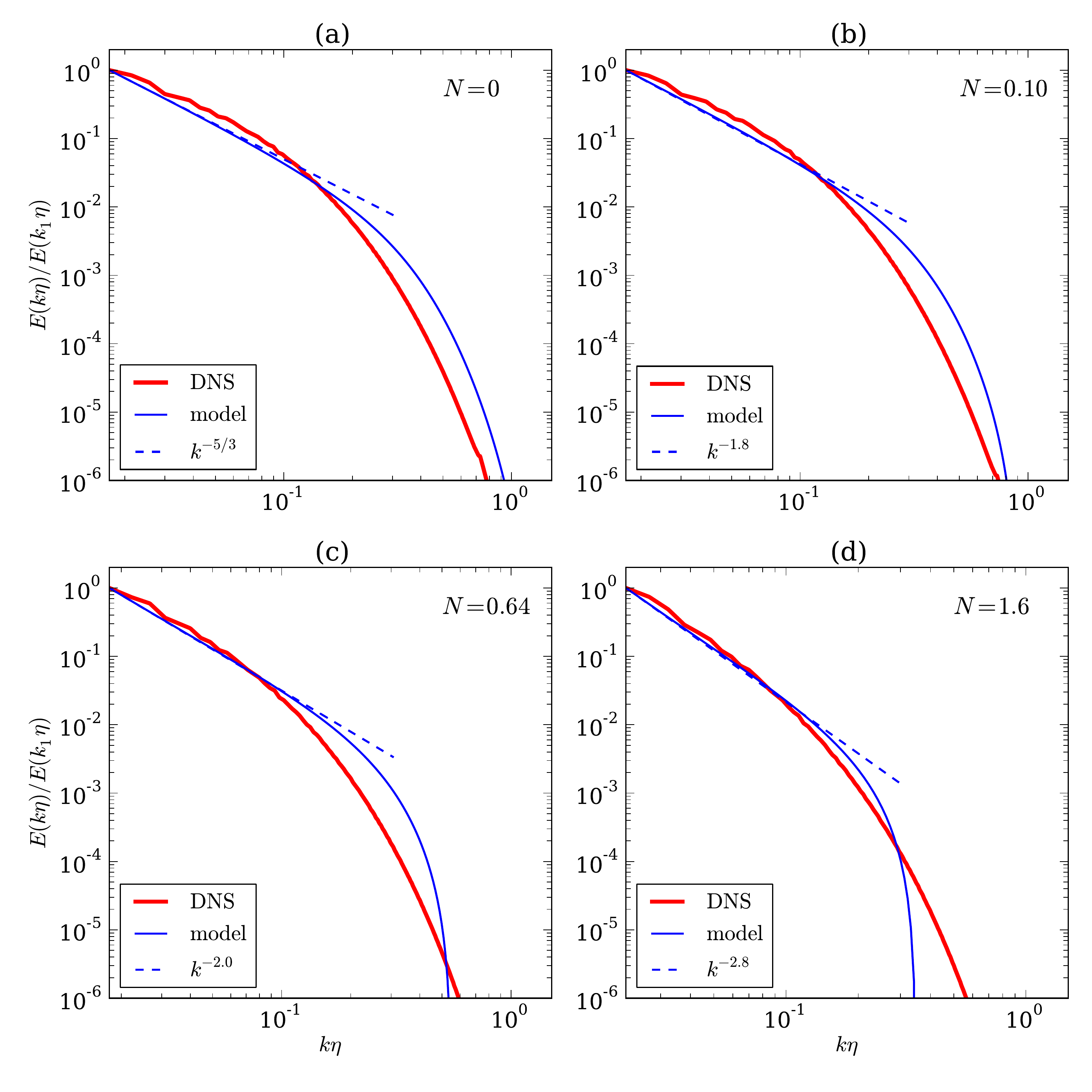}
\caption{ Plots of normalized energy spectra $E(k\eta)/E(k_1\eta)$ for: (a) $N=0$, (b) $N=0.10$, (c) $N=0.64$, and (d) $N=1.6$.  The dashed lines are the best fit curves.}
\label{fig:spectrum}
\end{center}
\end{figure}

For a small interaction parameter, we compute the model predictions for the normalized energy flux $\Pi(k)/\Pi_0$ and the normalized energy spectrum $E(k\eta)/E(k_1\eta)$ by substituting $N$ in   Eqs.~(\ref{eq:flux_exact}) and (\ref{eq:spectrum_exact}) respectively.   Since $N$ is small,  isotropic energy spectrum or $g(\theta) = 1$ is a good approximation, thus $c_1=1$ and $c_2=1/2$ (see Eqs.~(\ref{eq:c1_smallN}, \ref{eq:c2_smallN})).
In Figs.~\ref{fig:flux} and~\ref{fig:spectrum} we plot these quantities. 

To compare with the numerical results, we first compare the model predictions and numerical results for $N=0$, which corresponds to the pure fluid.  The numerical and model results, shown in Figs.~\ref{fig:flux}(a) and \ref{fig:spectrum}(a), match reasonably well, specially in the in inertial range; the energy flux is a constant, while the energy spectrum varies as $k^{-5/3}$.  This result  validates our model for the fluid turbulence.  

After this, we compare the numerical and model results for $N=0.10, 0.64, 1.6$; the energy fluxes and spectra are shown in Figs.~\ref{fig:flux}(b,c,d) and \ref{fig:spectrum}(b,c,d) respectively.  Figure~\ref{fig:flux} shows that for $N>0$, the energy flux is no more constant in the inertial range, and it decreases with $k$.  The model predictions and the numerical results are in a reasonable agreement with each other in the inertial range.  The deviations between the two results in the dissipative range indicates that the function $f_\eta(k\eta)$ of Eq.~(\ref{eq:Ek_smallN}) needs to be modified.  We attempted several alternatives, e.g., an exponential function, but they appear to perform worse.  A comprehensive work in this direction is required for a better agreement in the dissipative regime.

The energy spectrum shown in  Fig.~\ref{fig:spectrum} indicates that the energy spectrum gets steepened with the increase of $N$.  The spectral indices for $N=0.10, 0.64$ and 1.6 are $-1.8, -2.0$ and $-2.8$ respectively, which are steeper that Kolmogorov's $-5/3$ spectral index for hydrodynamic turbulence.  These results are in good agreement with earlier experimental\cite{Branover:PTR1994,Eckert:HFF2001} and  numerical works.\cite{Reddy:POF2014,Burattini:PF2008}

For interaction parameters far beyond unity, model $A$ is not valid because the energy spectrum tends to be anisotropic, and deviates from power law.  In the next subsection, we will employ model $B$ for large $N$, and compare the model predictions with  numerical results.

\subsection{Large interaction parameters}\label{subsec:large}
\begin{figure}[h]
\begin{center}
\includegraphics[scale=0.55]{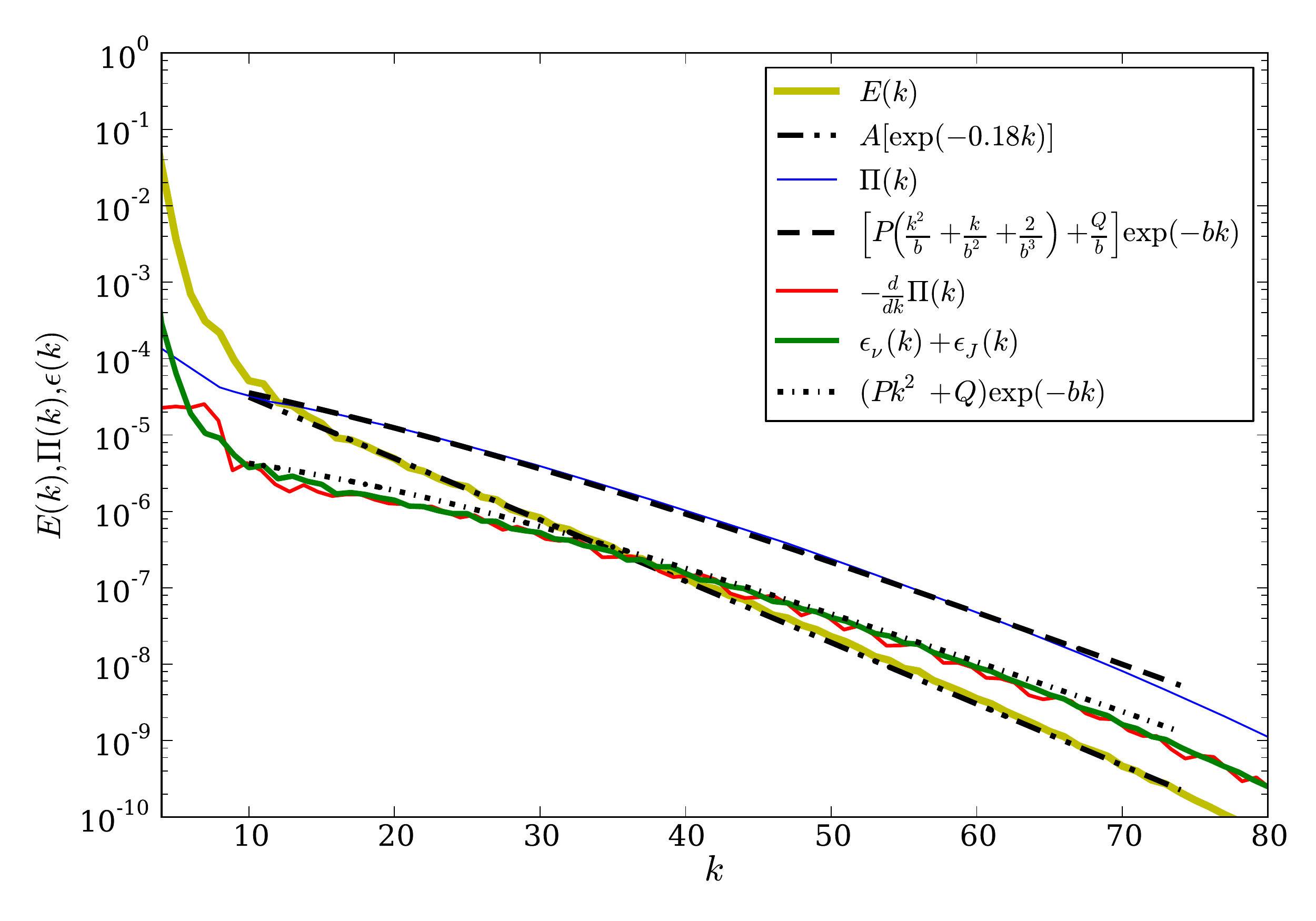}
\caption{For $N=130$, plots of kinetic energy spectrum $E(k)$, flux $\Pi(k)$, total dissipation $\epsilon(k) = \epsilon_J(k) + \epsilon_\nu(k)$, and $-\dfrac{d}{dk}\Pi(k)$. Note that $-\dfrac{d}{dk}\Pi(k) \approx \epsilon(k) $, consistent with Eq.~(\ref{eq:dPidk}).
The black double dot-dash,  dashed, dash-dot lines are the best fit curves for $E(k)$, $\Pi(k)$ and $\epsilon(k)$ respectively.}
\label{fig:N130}
\end{center}
\end{figure}

\begin{figure}[h]
\begin{center}
\includegraphics[scale=0.55]{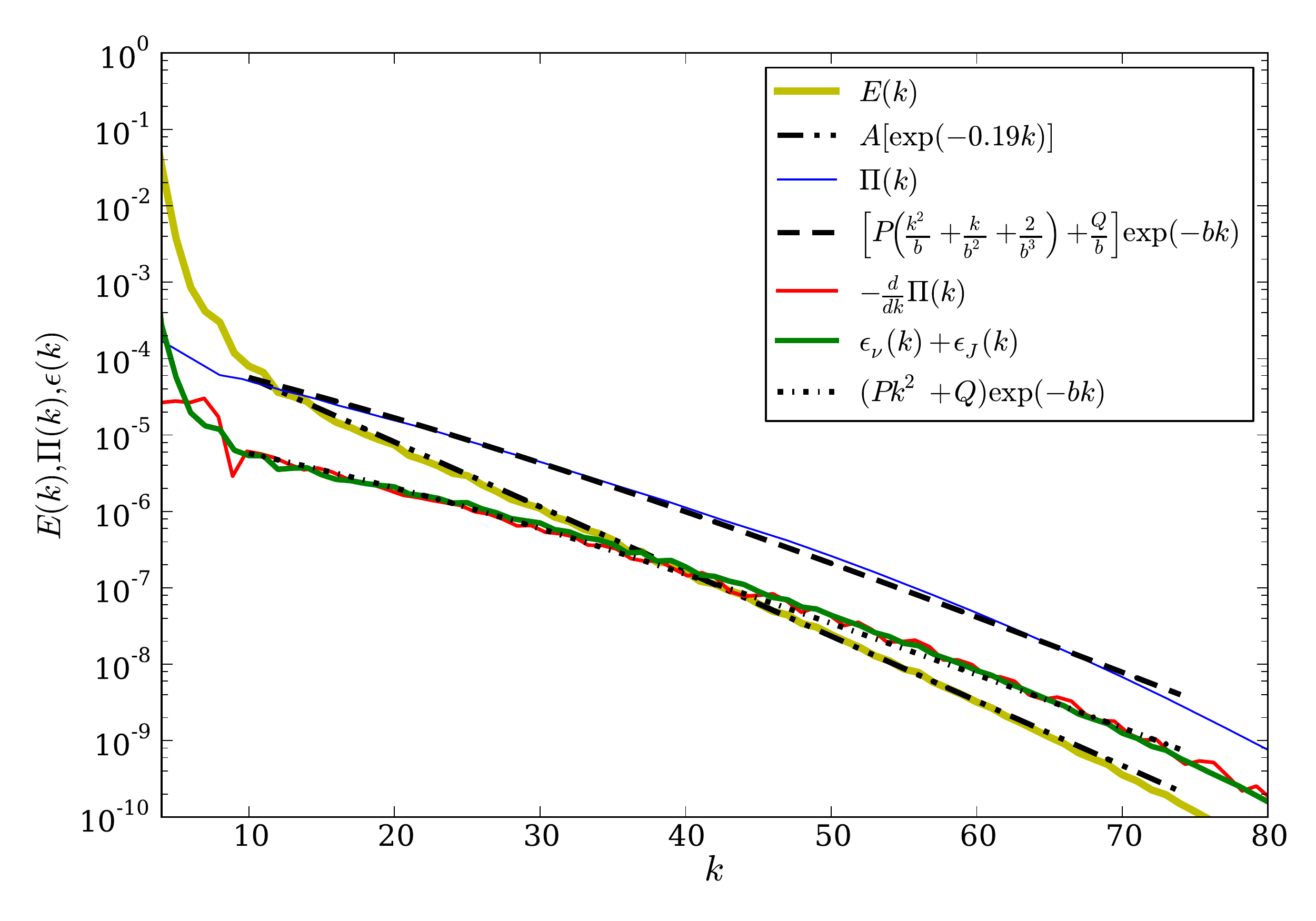}
\caption{For $N=220$, plots of kinetic energy spectrum $E(k)$, flux $\Pi(k)$, total dissipation $\epsilon(k) = \epsilon_J(k) + \epsilon_\nu(k)$, and $-\dfrac{d}{dk}\Pi(k)$. The black double dot-dash,  dashed, dash-dot lines are the best fit curves for $E(k)$, $\Pi(k)$ and $\epsilon(k)$ respectively.}

\label{fig:N220}
\end{center}
\end{figure}

We perform numerical simulations for $N=130$ and 220, and compute the energy spectra, dissipation spectra, and fluxes using the steady-state data.  These quantities are plotted in Figs.~\ref{fig:N130} and \ref{fig:N220} for $N=130$ and 220 respectively.   We fit the the numerical results with the expressions given by Eqs.~(\ref{eq:Ek_largeN}-\ref{eq:Pik_largeN}).  As shown in Figs.~\ref{fig:N130} and \ref{fig:N220}, the model predictions for the energy spectrum and energy flux  fit very well with the numerical results.      We also compute the ring spectrum $E(k,\theta)$, from which we compute $g(\theta)$ of Eq.~(\ref{eq:Ek}).

We compute the parameters $A, b, P$, and  $Q$ using the best fit curves for the energy and dissipation spectra.  These parameters are listed in Table~~\ref{tab:largeN}.  We also compute the constants $c_1$ and $c_2$ by substituting these parameter values  in the nondimensionalized form of Eqs.~(\ref{eq:largeN_P},\ref{eq:largeN_Q})
\begin{eqnarray}
P &=& 2Ac_1 \nu', \\
Q &=& 2Ac_2   B_0^{'2}, \label{eq:Q_nondim}
\label{eq:largeN_PQ_nondim}
\end{eqnarray}
and list them in Table~\ref{tab:largeN}.  We observe that $c_1 \approx 1$, consistent with Eq.~(\ref{eq:c1}), but $c_2$  differs significantly from 1/2,  indicating a strong anisotropy of the flow. We also compute $c_2$ by substituting numerically computed $g(\theta)$ in   Eq.~(\ref{eq:c2}).  The result, listed in Table \ref{tab:largeN} as $c_2'$, is within a factor of 3 of $c_2$ computed using Eq.~(\ref{eq:Q_nondim}).   Hence, the parameters are consistent with each other.

The aforementioned results shows that model $B$ describes the energy spectrum and flux for large $N$ quasi-static MHD very well.
 
\begin{table}[htbp]
\caption{The parameters of model $B$ defined in Eqs.~(\ref{eq:largeN_P},\ref{eq:largeN_Q}) computed using the simulation data.  The constants $c_1, c_2$ are computed using Eqs.~(\ref{eq:largeN_P},\ref{eq:largeN_Q}), while $c_2'$ is  obtained by substituting numerically computed $g(\theta)$ in Eq.~(\ref{eq:c2}).  }
\begin{ruledtabular}
\begin{tabular}{ccccccccc}
$N$ 	& $b$  & $A$	 & $P$	& $Q$ &	$c_1$ &$c_2$ &$c'_2$\\ 
\hline
$130$	& 0.18		& $2.0\times 10^{-4}$	 & $1.43 \times 10^{-7}$	& $1.7 \times 10^{-5}$   	& 0.99 & $6.8\times 10^{-5}$ & $1.4\times 10^{-4}$ \\
$220$	& 0.19		& $3.8\times 10^{-4}$	 & $2.51\times 10^{-7}$ & $3.8 \times 10^{-5}$ 	& 0.92 & $4.7\times 10^{-5}$ & $1.3\times 10^{-4}$ \\
\end{tabular}
\end{ruledtabular}
\label{tab:largeN}
\end{table}

\section{Conclusions}\label{sec:conclusions}

In this paper we present two models for quasi-static MHD.  The first model, which is applicable to small interaction parameters $N$, provides variable energy flux arising due to the Joule dissipation.  Consequently, the energy spectrum is steeper than that  of Kolmogorov's theory ($k^{-5/3}$).  The model predicts that the spectral index decreases with the increase of $N$.  The second model for very large interaction parameters predicts that the energy flux and spectrum are proportional to $\exp(-b k)$.  The model has several parameters that are determined by the numerical or experimental data. 

We validated our model predictions with numerical simulations.  We observe that the model results are in good agreement with the numerical results.  We compute the parameters of the second model using the numerical data.   Our models are also consistent with earlier numerical simulations\cite{Reddy:POF2014,Burattini:PD2008} and experimental results.\cite{Branover:PTR1994,Eckert:HFF2001}

Our models, based on variable energy flux, provides valuable insights into the physics of quasi-static MHD.  These models would be very useful for understanding experimental results and design of engineering applications.

\acknowledgments
 We are grateful to Mani Chandra for useful comments and help.  The computations were performed at the HPC system of IIT Kanpur.  This work was supported by a research grant  SERB/F/3279/2013-14 from Science and Engineering Research Board, India.

%

\end{document}